\documentclass[a4paper,10pt]{article}
\usepackage{amssymb}

\def\Eo{{\mathbb E}}

\def\Mo{{\mathbb M}}
\def\Qo{{\mathbb Q}}
\def\Ro{{\mathbb R}}
\def\Xo{{\mathbb X}}

\def\upd{{\rm d}}

\newcommand{\argmin}{\,{\rm arg\, min}\,}
\newcommand{\be}{\begin{eqnarray}}
\newcommand{\ee}{\end{eqnarray}}

\def\beginproof{\par\noindent{\bf Proof}\par}
\def\endproof{\par\strut\hfill$\square$\par\vskip 0.5cm}

\newtheorem{theorem}{Theorem}[section]

\newtheorem{proposition}[theorem]{Proposition}

\newenvironment{definition}[1][Definition]{\begin{trivlist}
\item[\hskip \labelsep {\bfseries #1}]}{\end{trivlist}}

\begin{document}
\title {The exponential family in\\ abstract information theory}
\author{Jan Naudts \and Ben Anthonis\\
University of Antwerp, Physics Department\\
Universiteitsplein 1, 2610 Wilrijk-Antwerpen, Belgium\\
{\tt jan.naudts@ua.ac.be, ben.anthonis@ua.ac.be}
}
\maketitle
\begin{abstract}
We introduce generalized notions of a divergence function and
a Fisher information matrix. We propose to generalize the notion of
an exponential family of models by reformulating it
in terms of the Fisher information matrix.
Our methods are those of information geometry.
The context is general enough to include applications from outside statistics. 

\end{abstract}

\section{Introduction}

The literature contains several generalizations
of the concept of models belonging to the exponential family \cite {BN78}.
See for instance \cite {NJ04,GD04,ES06,NJ09,AO11,PG12}.
The present work gives such a definition
in a context of an abstract information theory, which is not necessarily
based on probability. The main tools are those of information geometry
\cite{AN00}, in particular generalized divergence functions
\cite {NJ04,GD04,ES06,BLM67,CI75,AC10}.
They can be used to define a generalized Fisher information matrix
and generalized exponential families
(Definitions 1 and 2  in Section \ref {sect:def}).
 
The motivation for the present work comes from physics.
Applications of the new definitions in 
the context of classical and of quantum mechanics will follow in
a separate publication \cite {NA13}.
A preliminary write-up of the present work, including one non-statistical
example, is found in \cite {NA12}.

The next section introduces a generalized divergence in an abstract setting.
The Bregman divergence, discussed in Section \ref {sect:bregman}, is an important
subcase. Section \ref {sect:expfam} introduces our definitions of generalized
Fisher information and of generalized exponential families.
Sufficient conditions for a family to belong to a generalized family
follow in Section \ref {sect:suff}. The final two sections show
how our definitions relate to other generalizations found in the literature.

\section{Definitions}
\label{sect:def}

The abstract information framework $\Xo,\Mo,\Qo,\mu$ consists
of a topological space $\Xo$, a differentiable manifold $\Mo$,
and a linear space $\Qo$ of real functions of $\Xo$.
In addition there is given a continuous map $\mu:\,\Xo\rightarrow \Mo$.
The space $\Xo$ contains data sets. The map $\mu$ associates a
model point with each data set. The space $\Qo$ contains
questions about the data sets. To stress that $\Qo$ is not necessarily an algebra
the notation $\langle x|q\rangle$ is used rather than $q(x)$
to evaluate $q\in \Qo$ in the point $x\in\Xo$.
The constant function 1 belongs to $\Qo$ and satisfies $\langle x|1\rangle=1$
for all $x$ in $\Xo$.

A generalized divergence is a map $D:\,\Xo\times\Mo\rightarrow [0,\infty]$
satisfying the conditions
\begin{itemize}
 \item (compatibility) for each $x$ in $\Xo$ is $\mu(x)$
 the unique element of $\Mo$ minimizing the divergence $m\rightarrow D(x||m)$;
 \item (consistency) for each $m$ in $\Mo$ is $0=\inf_x\{D(x||m):\,\mu(x)=m\}$.
\end{itemize}
The divergence is interpreted as the amount of information which is lost when
the data set $x$ is replaced by the model point $m$.

Throughout the paper we assume that
there exist functions $\xi:\,\Mo\rightarrow \Ro$,
$\zeta:\Xo\rightarrow \Ro$ and a diffeomorphism $L:\,\Mo\rightarrow\Qo$ such that
for all $x\in\Xo$ and $m\in\Mo$ one has
\be
D(x||m)=\xi(m)-\zeta(x)-\langle x|L m\rangle.
\label{logmap:div}
\ee
From the compatibility condition follows the requirement that the map
$m\rightarrow \xi(m)-\langle x|L m\rangle$
is minimal when $m=\mu(x)$.
From the the positivity $D(x||m)\ge 0$ and the consistency condition follows
\be
\xi(m)&=&\sup_x\{\zeta(x)+\langle x|L m\rangle\}\cr
&=&\sup_x\{\zeta(x)+\langle x|L m\rangle:\,\mu(x)=m\}.
\label {logmap:cons}
\ee
The function $\zeta$ has the meaning of an entropy function.
The map $L$ is called the logarithmic map because in the standard case (see below)
it is essentially the natural logarithm.
The function $\xi$ is called the corrector \cite{TF07}.
We assume in what follows that it is a differentiable function.

\section{Bregman divergence}
\label {sect:bregman}

The obvious example of our framework is that of a statistical model.
Let $\Xo$ be the affine space of probability distributions over a finite alphabet $A$.
A question $q\in\Qo$ is a real function of $A$.
The evaluation of $q$ in the point $x$ is given by
\be
\langle x|q\rangle=\Eo_x q=\sum_{a\in A}x(a)q(a).
\ee
Let $\theta\in\Theta\subset\Ro^n\rightarrow m_\theta\in\Xo$
be a statistical model with sufficiently nice properties so that the set
\be
\Mo=\{m_\theta:\theta\in\Theta\}\subset\Xo
\ee
is a differentiable manifold.

A divergence of the Bregman type \cite {BLM67,AC10} is defined by
\be
D(x||m)&=&\sum_a\left[F(x(a))-F(m(a))-(x(a)-m(a))f(m(a))\right]\cr
&=&\sum_a\int^{x(a)}_{m(a)}\upd u\,\left[f(u)-f(m(a))\right],
\label{appl:bregman}
\ee
where $F$ is any strictly convex function defined on the interval $(0,1]$
and $f=F'$ is its derivative.
The standard case, involving the Boltzmann-Gibbs-Shannon entropy,
is recovered when $F(u)=u\ln u$.

Assume that the function $F$ is twice differentiable.
The logarithmic map is given by $Lm(a)=f(m(a))$.
The entropy function is $\zeta(x)=-\sum_a F(x(a))$. The consistency condition
(\ref  {logmap:cons}) follows from the convexity of the function $F(u)$.
Indeed, it implies that
\be
-F(x(a))\le -F(m(a))-(x(a)-m(a))f(m(a))
\ee
so that
\be
\zeta(x)+\langle x|Lm\rangle
&=&
\sum_a\left[-F(x(a))+x(a)f(m(a))\right]\cr
&\le&\sum_a\left[-F(m(a))+m(a)f(m(a))\right]\cr
&=&\zeta(m)+\langle m|Lm\rangle.
\ee
This implies (\ref  {logmap:cons}).
The model map $\mu$ is given by
\be
\mu(x)=\argmin_m\{\xi(m)-\langle x|L m\rangle\},
\ee
assuming existence and uniqueness of the minimum.

\section{Generalized exponential families}
\label{sect:expfam}

Introduce now coordinates $\theta\rightarrow m_\theta$ for the model manifold $\Mo$.
Use the notations $\xi(\theta)\equiv\xi(m_\theta)$ and $D(x||\theta)\equiv D(x||m_\theta)$.
By assumption the functions $\xi(\theta)$ and $\theta\rightarrow \langle x|Lm_\theta\rangle$
are differentiable. Therefore the first derivatives
\be
\frac {\partial\,}{\partial\theta^k}D(x||\theta)
\ee
vanish when $m_\theta=\mu(x)$.

\begin{definition}
The matrix of second derivatives
\be
I_{k,l}(x)
&=&\frac {\partial^2\,}{\partial\theta^k\partial\theta^l}
D(x||\theta)\bigg|_{m_\theta=\mu(x)}
\label {fisher:defg}
\ee
is the generalized Fisher information matrix.
\end{definition}

\begin{proposition}
The matrix $I_{k,l}(x)$
is covariant under coordinate transformations.
\end{proposition}
\beginproof
Let $\eta$ be a function of $\theta$. 
One calculates
\be
\frac {\partial^2\,}{\partial\theta^k\partial\theta^l}
D(x||\theta)
&=&\frac {\partial^2\,}{\partial\eta^m\partial\eta^n}D(x||\theta)
\frac {\partial\eta^m}{\partial\theta^k}
\frac {\partial\eta^n}{\partial\theta^l}\cr
& &
+\left(\frac {\partial\,}{\partial\eta^m}D(x||\theta) \right)
\frac {\partial^2\eta^m}{\partial\theta^k\partial\theta^l}.
\ee
The latter term vanishes when $m_\theta=\mu(x)$.
What remains is covariant under coordinate transformations.

\endproof

\begin{definition}
The model $\Xo,\Mo,\Qo,D$
belongs to a generalized exponential family if 
the Fisher information matrix $I_{k,l}(x)$, defined by (\ref {fisher:defg}),
is constant on the fibers ${\cal F}_m\equiv\{x:\,\mu(x)=m\}$.
\end{definition}
The constant value is then denoted $I_{k,l}(\theta)$.

A justification  of this definition follows later on
from the study of the definition in the familiar context of divergencies
of the Bregman type.
The main advantage of the above definition is that it does not specify
a particular choice of coordinates. That the Fisher information matrix
is constant on the fiber ${\cal F}_m$ is a scaling property. It means that
locally the manifold looks always the same,
independent of the point of view $x\in{\cal F}_m$.

\section{Sufficient conditions}
\label{sect:suff}

It is obvious to define a divergence between model points by
\be
D(m||n)=\inf_x\{D(x||n):\,\mu(x)=m\}.
\label {fisher:div}
\ee
It satisfies $D(m||n)\ge 0$.
Because of the special form (\ref {logmap:div}) of the divergence
there follows
\be
D(m||n)
&=&\xi(n)-\sup_x\{\zeta(x)+\langle x|Ln\rangle:\,\mu(x)=m\}.
\label {fish:ineq}
\ee
Using the consistency condition (\ref{logmap:cons}) one can write
\be
D(m||n)
&=&\sup_x\{\zeta(x)+\langle x|Ln\rangle:\,\mu(x)=n\}\cr
& &-\sup_x\{\zeta(x)+\langle x|Ln\rangle:\,\mu(x)=m\}.
\ee
In particular, $D(m||m)=0$ holds.

\begin{theorem}
Assume that the following Pythagorean relation$^{\rm \cite{CI75}}$ holds
\be
x\in {\cal F}_\theta\quad\Rightarrow\quad
D(x||\theta)+D(\theta||\eta)=D(x||\eta).
\label {geo:pyth}
\ee
Then the model belongs to the generalized exponential family.
\end{theorem}

\beginproof
From (\ref {geo:pyth}) follows
\be
I_{k,l}(x)=\frac{\partial^2\,}{\partial\eta^k\eta^l}\bigg\vert_{\eta=\theta}D(x||\eta)
=\frac{\partial^2\,}{\partial\eta^k\eta^l}\bigg\vert_{\eta=\theta}D(\theta||\eta).
\ee
This shows that $I_{k,l}(x)$ is constant along the fiber ${\cal F}_\theta$.

\endproof

The Pythagorean equality (\ref {geo:pyth}) expresses the intuition that the projection $\mu$
on the model manifold $\Mo$ is orthogonal.

\begin{theorem}
 If the logarithmic map is of the form
 \be
 Lm_\theta=-\alpha(\theta)-q_0-\eta^k(\theta)q_k
 \label{equiv:expfam}
 \ee
 with functions $\alpha$ and $\eta^k$, and questions $q_0,q_k$ in $\Qo$, 
 then the Pythagorean relation (\ref {geo:pyth}) is satisfied.
In particular, the model belongs to a generalized exponential family.
 
\end{theorem}

\beginproof

Introduce the abbreviation $\Phi=\xi+\alpha$.
From the definition of $\xi$ follows that
\be
\Phi(\theta)
=\sup_x\{\zeta(x)-\langle x|q_0\rangle-\eta^k(\theta)\langle x|q_k\rangle\}.
\ee
Hence $\Phi$ depends on $\theta$ only via the functions $\eta^k$.
In combination with
\be
D(x||\theta)=\Phi(\theta)-\zeta(x)
+\langle x|q_0\rangle+\eta^k(\theta)\langle x|q_k\rangle
\ee
and the assumption that for each $x$ there is a unique $\theta$ minimizing
$D(x||\theta)$ one concludes that the map $\theta\rightarrow\eta$ is invertible.
This observation is essential to conclude that $\langle x|q_k\rangle$
is constant along the fibers ${\cal F}_\theta$. One has indeed for $x\in {\cal F}_\theta$
\be
0&=&\frac{\partial\,}{\partial\theta^k}D(x||\theta)\cr
&=&\frac{\partial\eta^m}{\partial\theta^k}\left[
\frac{\partial\Phi}{\partial\eta^m}+\langle x|q_m\rangle
\right]
\ee
so that
\be
\langle x|q_m\rangle=-\frac{\partial\Phi}{\partial\eta^m}.
\ee

Now calculate, still assuming that $x\in {\cal F}_\theta$,
and using that $\langle x|q_k\rangle$ is constant along ${\cal F}_\theta$,
\be
D(\theta||\theta')
&=&\inf_x\{D(x||\theta'):\,x\in {\cal F}_\theta\}\cr
&=&\Phi(\theta')-\sup_x\{\zeta(x)
-\langle x|q_0\rangle-\eta^k(\theta)\langle x|q_k\rangle
:\,x\in {\cal F}_\theta\}\cr
&=&\Phi(\theta')-\sup_x\{\zeta(x)
-\langle x|q_0\rangle
:\,x\in {\cal F}_\theta\}-\eta^k(\theta)\langle x|q_k\rangle\cr
&=&\Phi(\theta')-\Phi(\theta)+(\eta^k(\theta')-\eta^k(\theta))\langle x|q_k\rangle\cr
&=&D(x||\theta')-D(x||\theta).
\ee
This shows the Pythagorean relation.
\endproof

\section{Justification}

We now return to Section \ref {sect:bregman} which
deals with the Bregman divergence.
In this context we give an explicit characterisation
of the generalized exponential family and show that it is
satisfied by the more common definition.

Taking derivatives of (\ref {appl:bregman}) yields
\be
\frac{\partial}{\partial\theta^k}D(x||\theta)
&=&-\sum_a\left[x(a)-m_\theta(a)\right]
\frac{\partial \,}{\partial\theta^k}
f(m_\theta(a)).
\ee
and, assuming $\mu(x)=m_\theta$,
\be
I_{k,l}(x)
&=&\sum_af'(m_\theta(a))
\frac{\partial m}{\partial\theta^k}(a)
\frac{\partial m}{\partial\theta^l}(a)\cr
&-&\sum_a\left[x(a)-m_\theta(a)\right]
\frac{\partial^2\,}{\partial\theta^k\partial\theta^l}
f(m_\theta(a)).
\ee
Independence of $x$ along ${\cal F}_\theta$ implies
\be
I_{k,l}(\theta)
&=&\sum_af'(m_\theta(a))
\frac{\partial m}{\partial\theta^k}(a)
\frac{\partial m}{\partial\theta^l}(a)
\label{equiv:fi}
\ee
and
\be
\sum_a\left[x(a)-m_\theta(a)\right]
\frac{\partial^2\,}{\partial\theta^k\partial\theta^l}
f(m_\theta(a))=0
\quad\mbox{for all }x\in {\cal F}_\theta.
\label{equiv:soe}
\ee
One concludes that the model belongs to the generalized exponential family
if the set of equations (\ref {equiv:soe}) holds for all $x$ satisfying
$x\in {\cal F}_\theta$ and the normalization condition $\sum_ax(a)=1$.
With $f'(u)=1/u$ expression (\ref{equiv:fi}) reduces to the standard expression
for the Fisher information matrix.

The obvious solution of (\ref {equiv:soe}) is that
there exist coordinates $\eta(\theta)$ such that
\be
\frac{\partial^2\,}{\partial\eta^k\partial\eta^l}
f(m_\theta(a))
\quad\mbox{ does not depend on }a.
\label{equiv:sol}
\ee
Indeed, (\ref{equiv:soe}) can be written as
\be
0&=&
\sum_a\left[x(a)-m_\theta(a)\right]
\left(
\frac{\partial^2\,}{\partial\eta^m\partial\eta^n}f(m_\theta(a))\right)
\frac{\partial\eta^m}{\partial\theta^k}
\frac{\partial\eta^n}{\partial\theta^l}\cr
& &
+\sum_a\left[x(a)-m_\theta(a)\right]
\left(\frac{\partial\,}{\partial\eta^m}f(m_\theta(a))\right)
\frac{\partial^2\eta^m}{\partial\theta^k\partial\theta^l}.
\ee
Because of the {\sl ansatz} (\ref{equiv:sol}) the former term vanishes.
The latter vanishes because $x\in {\cal F}_\theta$.

The requirement (\ref{equiv:sol}) is equivalent with the existence of
functions $q_0$ and $q_k$ such that for all $a$ and one fixed $b$
\be
f(m_\theta(a))-f(m_\theta(b))
&=&q_0(a)+\eta^kq_k(a).
\ee
One obtains
\be
\langle x|f(m_\theta)\rangle
&=&f(m_\theta(b))+\langle x|q_0\rangle +\eta^k\langle x|q_k\rangle.
\label {equiv:common}
\ee
This expression is of the form (\ref {equiv:expfam})
 (note that $f(m_\theta(a))=Lm_\theta(a)$).

\section{Discussion}

We propose to replace current definitions of generalized exponential families
by one formulated in terms of a generalized
Fisher information --- see Section \ref{sect:expfam}.
The new definition can be used in a more abstract setting of information theory,
one which does not necessarily rely on probability theory. 

The central tool of the present paper is an asymmetric divergence $D(x||m)$
between data sets $x$ and model points $m$. Divergences of this kind occur
in game theory --- see for instance Section 8 of \cite {GD04}.
They generalize the notion of a Bregman divergence \cite{BLM67}.

The notion of a generalized exponential family is usually formulated
directly in terms of the function $f$ appearing
in the generalized divergence by an 
expression similar to (\ref {equiv:common}).
We propose here to use the divergence in the first place
to define a generalized Fisher information matrix.
The latter is then used to define the generalized exponential families.

In \cite{NJ04} the function $f$, occurring in(\ref {equiv:common})
and defining the logarithmic map $L$ of (\ref {equiv:common}),
is assumed to be of the form
\be
f(u)=\int_1^u\upd v\,\frac{1}{\phi(v)},
\ee
with $\phi$ positive and increasing, and is called a deformed logarithm.
The $\phi$-deformed exponential family is then defined by an expression
of the form (\ref {equiv:expfam}). See also \cite{PG12}.
The special case with $\phi(v)=v^q$ is the $q$-deformed logarithm considered
in non-extensive statistical physics \cite{NJ09,TC10,NJ11}.
The corresponding exponential
families coincide with Amari's $\alpha$-families \cite{AO11,AN00}.

An alternative for the Bregman divergence is the U-divergence \cite {ES06}.
In our notations it reads
\be
D_U(x||m)=\sum_a\int_{f(x(a))}^{f(m(a))}\upd u\,[g(u)-x(a)],
\ee
where $U$ is a convex increasing function, $g=U'$ and $f$ is the inverse function of $g$
(note that $f$ is the deformed logarithm, $g$ the deformed exponential function
in the language of non-extensive statistical physics).
The $U$-model is then introduced in \cite {ES06}
as a generalization of the exponential model
and is defined by a relation of the form (\ref {equiv:expfam}).

\section*{}

\end{document}